
%
%
\magnification=\magstep1
\hsize=6.0truein
\vsize=8.5truein
\hoffset=.5cm
\voffset=.5cm
\baselineskip=12pt
\parskip=4pt plus 2pt
\parindent=24pt
\topskip=0pt
\leftskip=0pt
\rightskip=0pt
\def\br{\hfill \break}
\def\endpage{\par\vfil\penalty-10000 \vfilneg}
\def\half{{1\over 2}}

\font\small=cmr9
\def\sub#1{\vskip 12pt {\noindent #1} \vskip 8pt}
\def\pub#1{{\noindent #1} \vskip 8pt}
\def\exp{{\rm exp}\,}
\def\ln{{\rm log}\,}
\def\tr{{\rm tr}\,}

\def\spa{\check\lambda}
\def\spc{\hat\lambda}
\def\lam{\lambda}
\def\bra#1{\langle\, #1 \! \mid}
\def\ket#1{\mid \! #1 \,\rangle}
\def\norm#1#2{\langle\, #1 \! \mid \! #2 \,\rangle}
\def\av#1{\langle #1 \rangle}
\def\ep{\epsilon}
\def\der{\partial}
\def\Z{{\bf Z}}
\def\R{{\bf R}}
\def\sumn{\sum\nolimits}
\nopagenumbers
\
\vskip 8pt
\rightline{January 1992}
\rightline{VPI-IHEP-92/1}
\vskip 24pt
{\parindent=0pt
\centerline{DISCRETE AND CONTINUUM VIRASORO CONSTRAINTS}
\centerline{IN TWO-CUT HERMITIAN MATRIX MODELS}
\vskip 24pt
\centerline{Waichi OGURA\footnote{*}{\small
OGURA@VTVM2.BITNET\br
Address after April 1, 1992: YITP Uji Research Center,
Kyoto University, Uji 611, Japan.}}
\vskip 16pt
\centerline{\it Institute of High Energy Physics}
\centerline{\it Department of Physics}
\centerline{\it Virginia Polytechnic Institute and State University}
\centerline{\it Blacksburg, Va.~24061-0435, U.S.A.\null}}
\vskip 100pt
\centerline{ABSTRACT}
{\narrower\smallskip\parindent=0pt
Continuum Virasoro constraints in the two-cut hermitian matrix
models are derived from the discrete Ward identities by means of
the mapping from the $GL(\infty )$ Toda hierarchy to the nonlinear
Schr\"odinger (NLS) hierarchy. The invariance of the string equation
under the NLS flows is worked out.
Also the quantization of the integration constant $\alpha$ reported
by Hollowood et al.~is explained by the analyticity of the
continuum limit. \smallskip}
\vskip 40pt
\centerline{Submitted to Nucl.\ Phys.\ B}
\endpage
\footline={\hss\tenrm\folio\hss}
\pageno=1

\beginsection 1. Introduction

Two dimensional quantum gravity has been solved non-perturbatively
by taking the double-scaling limit of the matrix model -- the dual
model for the 2d discretized gravity. Here one has a parameter
$\kappa^2$ which determines the way this double limit is approached.
The potential is fine-tuned as the size of matrix gets large,
and the contributions from different 2d topologies may be obtained
as a power series in $\kappa^2$ with the power of $\kappa^{-1}$
being the Euler number of the Riemann surface.
Surprisingly, the scaling property (dependence on the cosmological
constant) of the $\kappa^2$ expansion agrees order by order with
the genus expansion in the 2d quantum gravity coupled with conformal
matters.\footnote{$^{\star}$}{\small
Exact equivalence has been proved at least on a torus [2].} This is
the reason why $\kappa$ is called the string coupling constant.
Furthermore, one can determine the scaling limit non-perturbatively
by solving a characteristic non-linear ordinary differential
equation (string equation) satisfied by the string susceptibility.

Even though one can find infinitely many critical points in the
variety of models and their potentials, which all have double
scaling limits, these continuum limits can be classified into
hierarchical universality classes due to the integrability
emerging at these critical points.
Actually integrability is there even at the discrete level.
If the size of matrix is finite, say $N$, the hermitian one matrix
model can be embedded into the $GL(N)$ Toda system, with variation
of the coupling constants of the matrix model forming the Toda flows
[3], so that the continuum integrability may be understood by the
scaling from the discrete integrability.

\sub{1.1~~$GL(N)$ Toda Hierarchy}

We will explain the Toda flow briefly by following refs.\ [1,3]
and Neuberger [4]. After integrating out the degrees of freedom
for the similarity transformations, one may evaluate the matrix
integral
$$
Z\equiv {1\over N!}\int\nolimits_{N\times N\,{\rm hermitian}}
d^{N^2}\phi~\exp[-\beta\,\tr V(\phi )]
=\norm {\rm vac}{\rm vac},
\eqno(1.1)
$$
in terms of the $N$-fermion vacuum
$$
\ket {\rm vac}={1\over\sqrt{N!}}\,\left|
\matrix{P_0(\lam_1)&\cdots&P_0(\lam_N)\cr
            \vdots &\ddots&     \vdots\cr
    P_{N-1}(\lam_1)&\cdots&P_{N-1}(\lam_N)\cr}
\right| ,
\eqno(1.2)
$$
where the $(n+1)^{th}$ state $P_n(\lam )$ is an orthogonal monic
polynomial of degree $n$ defined by
$$
\int_{-\infty}^{\infty}d\lam\, e^{-\beta V(\lam )}\,
P_n(\lam )\, P_m(\lam )=h_n\, \delta_{n,m}.
\eqno(1.3)
$$
Thus we can find $Z=\prod_{n=0}^{N-1}\, h_n$, and the vacuum
expectation value of $\tr\phi^k$ may be calculated as
$$
\av {{1\over N}\,\tr\phi^k}={1\over N}\,\sumn_{n=0}^{N-1}\,
\bra n\spc^k\ket n,
\eqno(1.4)
$$
where $\spc$ is the operator multiplying with $\lam$, and $\ket n$
is the orthonormal basis defined by $\ket n=\exp (-{\beta\over 2}\,
V(\lam))\, P_n(\lam )/\sqrt{h_n}$.
Henceforth the r.h.s.\ will be simply denoted by $\av {\spc^k}$.

Though the potential $V(\lam )=\sumn_{k=0}^{\infty}\, g^k\,
\lam^k$ should be fixed at one of the critical points at the
double-scaling limit, at finite $N$ we are free to alter it from
the critical point, and find that
$$
\delta\, (P_n/\sqrt{h_n})=\sumn_{m=0}^{\infty}\,
O_{nm}\, P_m/\sqrt{h_m},
\eqno(1.5)
$$
where all upper half components of $O$ are empty. It is convenient
to work on the orthonormal basis, and define $Q_{nm}=\bra n\spc
\ket m$. $Q$ is a symmetric matrix, and $\delta V(Q)=O+\, ^t\!\, O$.
Thus $O$ can be identified with $[\delta V(Q)]_-$, where $[~~]_{\pm}$
denotes the projection such as $\half\, [~~]_{\rm diagonal}+
[~~]_{\rm upper/lower~triangular}$. Hence the variation of $Q$
can be obtained as
$$
\delta Q=[Q,\half\,\left( [\delta V(Q)]_+-[\delta V(Q)]_-\right) ].
\eqno(1.6)
$$

Toda Hamiltonian flows defined by Adler [3] are simply
$$
{\der Q\over\der g_k}=[Q,[Q^k]_+]=-[Q,[Q^k]_-],
\eqno(1.7)
$$
and from $\delta V(\lam)=V(\lam -\delta\lam )-V(\lam )$,
one also finds the discrete string equation
$$
[P,Q]=1,
\eqno(1.8)
$$
with $P=\half ([V'(Q)]_+-[V'(Q)]_-)$.
As proved by Martinec [3], the Toda flows are commutative,
and also preserve the string equation.

In terms of the Ward identities under the Toda flows,
Mironov and Morozov [5] have found the Virasoro constraints
$$
L_n\, Z=0,~~~~~~~(n\geq -1)
\eqno(1.9)
$$
where
$$
L_n=\sumn_{l=1}^{\infty}l\, g_l\,
{\der\over \der g_{l+n}}+\beta^{-2}\,\sumn_{l=0}^n
{\partial^2\over \der g_l\,\der g_{n-l}},
\eqno (1.10)
$$
satisfies the Virasoro algebra $[L_n,L_m]=(n-m)\, L_{n+m}$ for
$n,~m\geq -1$.

\sub{1.2~~Double-Scaling Limit}

The $GL(\infty )$ Toda hierarchy is transmuted into the KdV hierarchy
at the double-scaling limit, and more generally, it has been shown
by Douglas [6] that there appears the generalized KdV hierarchy
in the one-cut hermitian multi-matrix models.
The degree $n$ of the orthogonal polynomials may be described in
terms of a continuous parameter $t$ at the double-scaling limit,
such that $n/\beta =1-at$.\footnote{$^{\star}$}{\small
Note that $1-N/\beta$ is the cosmological constant.}
Then the pair $Q$ and $P$, which are the matrices defined similarly
for the first piece of the matrix chain, scale to the scalar Lax
pair [7] as $\beta\rightarrow\infty$ and $a\rightarrow 0$
simultaneously. In a critical multi-matrix model, for instance,
one can find the Lax operator $L=\der_t^q+u_{q-2}\,\der^{q-2}_t+
\ldots +u_0$, and (1.8) scales to the continuum string equation
$$
[(L^{p/q})_+,L]=1,
\eqno(1.11)
$$
where $u_{q-2}$,\dots ,$u_0$ are functions of $t$ called the scaling
functions, and $(~~)_+$ denotes the ordinary differential operator
part of the pseudo differential operator $L^{p/q}$.
Similarly to the Toda hierarchy, the generalized KdV hierarchy
has infinite number of commutative Hamiltonian flows
$$
{\der L\over\der t_p}=[(L^{p/q})_+,L]=-[(L^{p/q})_-,L],
\eqno(1.12)
$$
and in the matrix model, these generalized KdV flows can be
generated by the scaling operators $\sigma_k({\cal O}_{\alpha})$
with $p=kq+\alpha +1$ such that
$$
{\der F\over\der t_{k,\alpha}}
=\langle\sigma_k({\cal O}_{\alpha})\rangle
=\int_{\infty}^tdt\, \bra tL^{p/q} \ket t,
\eqno(1.13)
$$
where $F=\ln Z$, and the residue $H_{k,\alpha}=\bra tL^{p/q} \ket t$
is a differential polynomial of the scaling functions satisfying
certain recursion relations [8-10].
Integrating the string equation (1.11), one also finds the
Virasoro and the W$_n$-constraints $(n=3,\ldots ,q)$ on $Z$,
and thereby $Z$ can be identified with the $\tau$-function of the
generalized KdV hierarchy [9-11].
Since the integrability appearing in the continuum is merely a
consequence of the integrability in the discrete, it has been
conjectured that the Virasoro and the W-constraints can be
derived from the discrete Ward identities at any scaling limit.
This has been proved at least for the one-cut hermitian one
matrix models [11,12].

These constraints have been already found in the 2d topological
gravity coupled with the ADE series of the topological minimal
models [8,13], and very recently, the same Virasoro constraints
has also arisen in the Kontsevich model [14,15].

\endpage
\sub{1.3~~Overview}

The one-cut family of the hermitian matrix models has
demonstrated their rich structures in the generalized KdV
hierarchy and W-constraints.
In this paper, we shall study the two-cut family of the hermitian
one matrix models [16-18], and look for similar structures.
The procedure developed in ref.\ [12], which is easier to handle
and more straightforward than the one used in ref.\ [11], will be
applied to the two-cut family; and consequently, the continuum
Virasoro constraints will be derived from the discrete Ward
identities ($\S3$).
The result agrees with refs.\ [17,18], and the meaning of the
additional parameter $\alpha$ appearing in the continuum Virasoro
constraints will be clarified.
Eventually one can find the nonlinear Schr\"odinger (NLS) hierarchy
for the $2 \times 2$ matrix Lax operators [19,20], and obtain the
mKdV hierarchy as the reduction into the even potential models
[17,18]. The invariance of the string equation under the NLS flows
will be proved in $\S4$. Hollowood et al.~[18] have obtained what
they call the Zakharov-Shabat (ZS) hierarchy by rotating the
anti-hermitian model to the hermitian, and from the ZS hierarchy
one can get both KdV and mKdV hierarchies by reduction.
We will investigate this rotation further in $\S5.1$ according to
the mapping we found, and construct a rotation from the NLS
hierarchy to the ZS hierarchy.
Finally the quantization of $\alpha$ reported in ref.\ [18] will be
interpreted as the analyticity of the model at the continuum limit
$\ep\rightarrow 0$ by means of the rescaling of $\ep$ ($\S5.2$).

Generally one can show that the family of matrix models is
governed by two kinds of distinct integrable structures in the
discrete and the continuum, and the investigation on the relation
between these two structures is not just of mathematical interest,
but is essential in order to understand the physical outcome from
the integrability in the continuum.
As we will examine later, in order to reach a general point
$(t_1,t_2,\ldots )$, we must perturb the matrix model by the
associated flows, which may or may not be allowed within the
original model.
Especially the flow connecting two different critical points
must deviate from the matrix model.
This is because even though the critical points are individually
realized by the critical matrix models, in between one cannot find
the corresponding matrix models, in other words, if we impose the
physical conditions on the solution of the string equation in order
to extract the physics from the formal system like the generalized
KdV, the NLS, or the ZS hierarchy, those solutions generally become
unstable under the associated flows [21].
Unfortunately the mapping itself is not strong enough to solve
this reduction problem, but it will provide a foundation
to get further insight.

\beginsection 2. Two-cut Models

\pub{2.1~~Hierarchical Criticality}

We will determine the critical potentials of the hermitian
one matrix models, and classify them into hierarchies.
The string equation (1.8) governs the matrix model completely,
but here we use an alternative, but equivalent pair of equations
$$
\eqalignno{
\bra {n-1}V'(\spa )\ket n&={n\over\beta},&(2.1a)\cr
\bra nV'(\spa )\ket n&=0,&(2.1b)\cr}
$$
which we also call {\it the string equation}.
In terms of the recursion relation
$$
\lam P_n(\lam )=P_{n+1}(\lam )+S_n\, P_n(\lam )+R_n\, P_{n-1}(\lam ),
\eqno(2.2)
$$
one can show that
$$
\spa ~{\buildrel \rm def \over =}~
{1\over\sqrt{\hat h}}\,\spc\,\sqrt{\hat h}=
\hat z+\hat S+\hat z^{\dagger}\,\hat R,
\eqno(2.3)
$$
where $\hat R\ket n=R_n\ket n$, $\hat S\ket n=S_n\ket n$,
$\hat h\ket n=h_n\ket n$, $\hat z\ket n=\,\ket {n+1}$, and
$\hat z^{\dagger}\ket n=\,\ket {n-1}$.
Due to the hermiticity of $\spc$, (2.1a) may be rewritten as
$$
\bra n 1-\half\,\spa V'(\spa )\ket n=1-{n+\half\over\beta},
\eqno(2.1c)
$$
which is more convenient for our present purpose.
Now assume that $R_n$ and $S_n$ converge as $n\rightarrow\beta$,
and denote those limits by $a^2$ and $b$, respectively (the
positivity of $R_n$ follows from $R_n=h_n/h_{n-1}$), then (2.1c)
and (2.1b) may be evaluated by the contour integrals around $z=0$
$$
\eqalign{
&\int_{C_0}{dz\over z}\left[
1-\half\lam (z)\, V'(\lam (z))\right]=0,\cr
&\int_{C_0}{dz\over z}\, V'(\lam (z))=0,\cr}
\eqno(2.4)
$$
with $\lam (z)=z+b+a^2/z$. We can rewrite (2.4) further in
terms of $z=[\lam-b-\sqrt{(\lam -b)^2-4a^2}\, ]/2$ such that
$$
\eqalign{
\int_{C_{\infty}}d\lam\,\left[
(\lam -b)^2-4a^2\right]^{-\half}\,&\left[ 1-\half\lam\,
V'(\lam )\right] =0,\cr
\int_{C_{\infty}}d\lam\,\left[
(\lam -b)^2-4a^2\right]^{-\half}\,&V'(\lam )=0,\cr}
\eqno(2.5)
$$
where the contour $C_{\infty}$ must be large enough to enclose
all singularities on the complex plane.
Noticing that
$$
\int_{C_{\infty}}d\eta\, (\eta^2-4a^2)^{-\half}\,\eta^m=0,
\eqno(2.6)
$$
for $m$ odd integral or negative even integral,
it is easy to obtain a general solution
$$
\eqalign{
1-\half\lam\, V'(\lam )&=\left[ \lam F(\lam )\right]_+,\cr
F(\lam )&=c(\lam )\, (\lam -\lam_+)^{\half}\,
(\lam -\lam_-)^{\half}.\cr}
\eqno(2.7)
$$
Here $\lam_{\pm}=b\,\pm\, 2a$, and $[~~]_+$ denotes the principal
part (including the $\lam^0$ term) of the Laurent expansion
about $\lam =\infty$. $c(\lam )$ is an arbitrary entire function
satisfying Res$_{\lam =\infty}\, F(\lam )=-1$ in order that the
$\lam^0$ term of $\lam F(\lam )$ equals $1$. Note that the first
line in (2.7) may be expressed as $V'(\lam )=-2\left[ F(\lam
)\right]_+$. $F(\lam )$ controls the eigenvalue distribution of
$\spc$ through (see {\it e.g.} [22])
$$
\rho(\lam )={1\over\pi}\,\lim_{\ep\to 0+}{\rm Im}\, F(\lam -i\ep ).
\eqno(2.8)
$$
Hence $c(\lam )$ must be negative along the support of $\rho(\lam )$,
which turns out to be the interval $(\lam_-,\lam_+)$ separated by the
even order zeros of $c(\lam)$ (no odd order zeros are allowed along
$(\lam_-,\lam_+)$). One might think $c(\lam )$ could have cuts along
$(\lam_-,\lam_+)$, but this cannot happen because of (2.5) and the
negativity of $c(\lam )$.
We will see, for the critical potential $V(\lam )$ given by (2.7),
that (2.1a) and (2.1b) scale to the continuum string equation at
the double-scaling limit, and this scaling behavior is completely
governed by the eigenvalue distribution at the edge, {\it i.e.}
the zeros of $F(\lam )$.

Suppose $c(\lam )$ has a zero only at $\lam =\lam_+$,
then $c(\lam )$ may be expanded as
$$
c(\lam )=\sumn_{k\geq 1}\,c_k\, (\lam_+-\lam )^k,
\eqno(2.9)
$$
where $c_k$ must satisfy
$\sumn_{k\geq 1}\, (2a)^{k+2}\,{(2k+1)!!\over (k+2)!}\, c_k=-1$.
This is the family of one-cut models, in which the $m^{th}$
critical potentials can be obtained by choosing $c_m<0$ and
$c_k=0$ for $k<m$, where the negativity of $c_m$ follows from
the negativity of $c(\lam )$. More generally, $c(\lam )$
may have zeros at both ends such as
$$
c(\lam )=\sumn_{k_+,k_-\geq 0}\, c_{k_+,k_-}\,
(\lam_+-\lam)^{k_+}\, (\lam -\lam_-)^{k_-},
\eqno(2.10)
$$
which is the doubling studied in ref.\ [16]; in particular,
$k_+=k_-$ gives the family of even one-cut models.
All of these one-cut families yield the KdV hierarchy at the
double-scaling limit, whereas the mKdV hierarchy and more generally
the NLS hierarchy may be obtained from the two-cut family as follows.

Suppose $c(\lam )$ has an even order zero at $\lam =b$
$$
c(\lam )=\sumn_{k\geq 1}\, c_{2k}\,(\lam -b)^{2k},
\eqno(2.11)
$$
then $c_{2k}$ must satisfy $\sumn_{k\geq 1}\,
(2a^2)^{k+1}\,{(2k-1)!!\over (k+1)!}\, c_{2k}=-1$.
These are the two-cut models studied in ref.\ [18], in which
the $2m^{th}$ critical potentials\footnote{$^{\star}$}{\small No
odd critical potentials are allowed within the hermitian models.}
are given by $c_{2m}<0$ and $c_{2k}=0$ for $k<m$; in particular
we call $c_{2k}=-{(m+1)!\over (2a^2)^{m+1}\, (2m-1)!!}\,
\delta_{k,m}$ the $2m^{th}$ critical point.
More general critical model can be obtained by taking the product
of these two kinds of critical models in the same manner as the
$(k_+,k_-)$ model has been constructed from the doubling of the
${k_+}^{th}$ and ${k_-}^{th}$ critical one-cut models in ref.\ [16].

\sub{2.2~~String Equation}

We will fix our notation for the two-cut models by following
Crnkovi\'c and Moore [17], and give a heuristic argument to derive
the continuum string equations. The $2m^{th}$ double-scaling limit
may be defined by $\ep\rightarrow 0$ in
$$
\eqalign{
R_{n}&=1+(-1)^n\, 2\ep\, f(t)+\ldots ,\cr
S_{n}&=b+(-1)^n\, 2\ep\, g(t)+\ldots ,\cr}
\eqno(2.12)
$$
with $x=n/\beta =1-\ep^{2m}\, t$, and $\kappa^2\,
\beta\,\ep^{2m+1}=1$ ($a$ has been normalized to $1$).
For the orthogonal two component vector defined by
$$
\eqalign{
P^+_{2l}(\lam )&=\exp (-{\beta\over 2}\, V(\lam))\, (-1)^l\,
P_{2l}(\lam )/\sqrt{h_{2l}},\cr
P^-_{2l+1}(\lam )&=\exp (-{\beta\over 2}\, V(\lam))\, (-1)^l\,
P_{2l+1}(\lam )/\sqrt{h_{2l+1}},\cr}
\eqno(2.13)
$$
$\spa$ may be represented as a $2\times 2$ matrix operator
$$
\spa \left(\matrix{P^+_{2l}\cr P^-_{2l+1}\cr}\right)=
\left(\matrix{\hat S & 1 -\,\hat z^{\dagger}\,\hat R\,
\hat z^{\dagger}\cr -\hat z^2+\hat R & \hat S\cr}\right)\,
\left(\matrix{P^+_{2l}\cr P^-_{2l+1}\cr}\right) .
\eqno(2.14)
$$
At the scaling limit, $P^{\pm}_n(\lam )$ scales to $\kappa\,
\sqrt{\ep}\,{\norm t{\lam}}_{\pm}$ with $\ket {\lam}$ being
a two-component vector, while both $\spa$ and $\spc$ scale as
$$
\bra t \spa \ket{\lam}=\bra tb\,E+2\,\ep\, L \ket {\lam}.
\eqno(2.15)
$$
Hence we will not distinguish $\spc$ from $\spa$. $E$ is the
identity $2\times 2$ matrix, and $L$ is the Lax operator
$L=-i\,\sigma_2\,\kappa^2\,\der_t-\sigma_1f+\sigma_3g$ with
$\sigma_i$'s being the Pauli matrices.

For the critical potential given by (2.11), the leading part
of (2.1a) is trivially satisfied due to (2.4), and the rest
of (2.1a) and (2.1b) can be combined into the string equation
$$
\bra tV'(\spc)\ket t=-i\,\ep^{2m} t\,\norm tt,
\eqno(2.16)
$$
where the factor $-i$ is necessary to compensate for the extra
$i$ factor contained in $\norm tt$ (see $\S2.3$). Based on the
same argument as ref.\ [12], all $\bra t\spc^n\ket t$ with $n$
negative vanish. Basically this is because of the analytic
continuation from $n$ positive to negative. Thus we can remove
the restriction $[~~]_+$ from $V'(\spc )$, and find that
$$
t\,\bra tL^0\ket t+4^{m+1}\, c_{2m}\,\bra tL^{2m}\ket t=0.
\eqno(2.17)
$$
This holds for any $2m^{th}$ critical potential, since the higher
order terms disappear as $\ep\rightarrow 0$.

\sub{2.3~~Diagonal Kernel of the Resolvent}

The explicit form of the residues for the general matrix Lax
operator $L$ can be obtained from the resolvent $L-\lam\, E$,
which has been studied in ref.\ [20] and their theorem 6 provides
us with the following results. The Laurent expansion of the diagonal
kernel of the resolvent about $\lam =\infty$ starts from the
zero-th order, and for $k\geq -1$
$$
\bra tL^k\ket t={i\over 2}\,\left( F_k\,\sigma_3 +
G_k\,\sigma_1+H_k\,E\,\right) ,
\eqno(2.18)
$$
where
$$
\eqalign{
F_{k+1}&=g\, H_k-\half\, G'_k,\cr
G_{k+1}&=-f\, H_k+\half\, F'_k,\cr
\half\, H'_k&=f\, F_k+g\, G_k,\cr}
\eqno(2.19)
$$
with $'$ denoting $D=\kappa^2\, \der_t$.
First two levels can be determined as $F_{-1}=G_{-1}=0$,
$H_{-1}=1$, $F_0=g$, $G_0=-f$, $H_0=0$ by assuming that $f$ and $g$
are $t$ independent and directly evaluating the diagonal kernel as
$$
\eqalign{
\bra t {1\over L-\lam\, E}\ket t&=\lim_{\ep\to 0+}\,
\int_{-\infty}^{+\infty}{dp\over 2\pi}\,
\left(\matrix{g-\lam +i\ep & -f-ip \cr
                 -f+ip & -g-\lam +i\ep \cr} \right)^{-1}, \cr
&={i\over 2\sqrt{\lam^2-f^2-g^2}}\,
\left(\matrix{-g-\lam & f   \cr
               f   & g-\lam \cr} \right) . \cr}
\eqno(2.20)
$$
Higher levels may be obtained from (2.19), where the first few are
$$
\eqalign{
F_1&=\half f',\cr
F_2&=-{1\over 4}\, g''+\half\, g(f^2+g^2),\cr} ~~~~
\eqalign{
G_1&=\half g',\cr
G_2&={1\over 4}\, f''-\half\, f(f^2+g^2),\cr} ~~~~
\eqalign{
H_1&=\half\, (f^2+g^2),\cr
H_2&=\half\, (gf'-fg'),\cr}
$$
$$
\eqalign{
F_3&=-{1\over 8}\, f'''+{3\over 4}f'(f^2+g^2),~~~~~~
G_3=-{1\over 8}\, g'''+{3\over 4}g'(f^2+g^2), \cr
H_3&=-{1\over 4}\, f\, f''+{1\over 8}\, (f')^2
     -{1\over 4}\, g\, g''+{1\over 8}\, (g')^2
     +{3\over 8}(f^2+g^2)^2 .\cr}
\eqno(2.21)
$$

\sub{2.4~~Reduction to the mKdV Hierarchy}

For an even potential model, one finds $S_n=0$ and therefore
$g=0$, for which (2.19) yields $F_{2k+1}=\half\, S'_k[f]$,
$G_{2k}=-S_k[f]$, $H_{2k+1}=R_{k+1}[u]-\half\, S'_k[f]=
R_{k+1}[\bar u]+\half\, S'_k[f]$, and otherwise zero,
where $u$ and $\bar u$ are defined by the Miura mapping
$u=f^2+f'$ and $\bar u=f^2-f'$, and $S_k[f]$ and $R_k[u]$ are
the $k^{th}$ Gel'fand-Dikii differential polynomial in the
mKdV and the KdV hierarchies, respectively. The definitions are
$$
\eqalign{
S_k[f]&=[(f-\half\, D)\, D^{-1}\, (f+\half\, D)\, D]^k\, f
=[-{1\over 4}\, D^2+f\, D^{-1}\, f\, D]^k\, f,\cr
R_k[u]&=[D^{-1}\, (f+\half\, D)\, D\, (f-\half\, D)]^k\, 1
=[-{1\over 4}\, D^2+\half\, u+\half\, D^{-1}\, u\, D]^k\, 1.\cr}
\eqno(2.22)
$$
The $2m^{th}$ string equations in the NLS hierarchy are therefore
$$
\eqalignno{
tg+4^{m+1}\, c_{2m}\, F_{2m}&=0,&(2.23a)\cr
tf-4^{m+1}\, c_{2m}\, G_{2m}&=0,&(2.23b)\cr}
$$
with $c_{2m}<0$.
(2.17) also provides the third string equation $H_{2m}=0$, but
since $H'_{2m}=0$ follows from (2.23), and $H_{2m}$ is odd under
either $f\rightarrow -f$ or $g\rightarrow -g$ while any of
$(\pm f,\pm g)$ satisfies (2.23), the third string equation may be
obtained by integrating (2.23).\footnote{$^{\star}$}{\small
The integration constant does not vanish if the odd modes are
mixed in.}

We will conclude this section with a few comments.
Since the scaling behavior does not depend on the parameter $b$,
we omit $b$ hereafter, while keeping $g$ general.
In the sphere limit $\kappa =0$, or as $t\rightarrow\infty$,
all $t$ derivative disappear, so that (2.20) provides exact
relations, such as
$F_{2m}={(2m-1)!!\over 2^{m}\, m!}\, g\, (f^2+g^2)^m$,
$G_{2m}=-{(2m-1)!!\over 2^{m}\, m!}\, f\, (f^2+g^2)^m$.
The string equation at the $2m^{th}$ critical point is thus given by
$$
t=2\, (m+1)\, (f^2+g^2)^m.
\eqno(2.24)
$$
For an even potential, the system is reduced by $g=0$, and only
(2.23b) remains nontrivial, which may be written down as
$$
tf={2^{m+1}\, (m+1)!\over (2m-1)!!}\, S_m[f],
\eqno(2.25)
$$
at the $2m^{th}$ critical point.
The same string equation has been found also in the one-cut family
of the unitary matrix models in ref.\ [23]; furthermore it is proved
that (2.25) has a unique, real, pole-free, solution consistent with
the above asymptotic behavior (2.24) for $t\rightarrow\infty$ [24].

\beginsection 3. Discrete and Continuum

\pub{3.1~~Mapping from Toda to NLS}

The partition function (1.1) is a function of coupling constants
through the potential $V(z)=\sumn_{k=0}^{\infty}\, g_k\, z^k$,
and satisfies the differential equation
$$
J(z)\, Z=\left[ \,\half\, V'(z)
-\langle\hat W(z)\rangle\, \right] Z.
\eqno (3.1)
$$
Here $\av{\hat W(z)}=\av{\beta^{-1}\,\tr [\, (z-\phi)^{-1}\,
]}$ is the generating function, and $J(z)=\sumn_{k\in\Z}$
$z^{-k-1} J_k$ is defined by $J_k =\beta^{-2}\, \der /\der
g_k$ for $k\geq 0$, and $J_{-k} ={k \over 2}\, g_k$ for
$k\geq 1$ [12], so that $J_k$ satisfies
$$
\left[\, J_k\, ,\, J_l,\right] =\half\,\beta^{-2}
k\,\delta_{k+l,0}.
\eqno(3.2)
$$
Re-expanding $J(z)$ as
$J(z)=\sumn_{k\in\Z}\, z^{-k}\, (z^2-4)^{-\half}\, \tilde J_k$,
one can find that
$$
\eqalign{
\tilde J_{2k}&=\int_{C_{\infty}}{dz\over 2\pi i}\,
(z^2-4)^{\half}\, z^{2k-1}\, J^{even}(z),\cr
\tilde J_{2k+1}&=\int_{C_{\infty}}{dz\over 2\pi i}\,
(z^2-4)^{\half}\, z^{2k}\, J^{odd}(z),\cr}
\eqno(3.3)
$$
where $J(z)$ is divided into odd and even functions of $z$,
denoted as $J^{even}(z)$ and $J^{odd}(z)$, respectively.
Note that the superscripts of $J$ are named according to the
even/odd nature of the potential, not its derivative.
$J^{odd}(z)$ disappears from $\tilde J_{2k}$ because of the
cancellation along $C_{\infty}$, and so does $J^{even}(z)$ from
$\tilde J_{2k}$. From (3.2) and (3.3), the commutation relations
for $\tilde J_k$ can be calculated as
$$
\left[\,\tilde J_k\, ,\,\tilde J_l\,\right] =\beta^{-2}\left[
\half\, k\,\delta_{k+l,0}-2\, (k-1)\,\delta_{k+l,2} \right] .
\eqno(3.4)
$$

According to (2.11), $V'$ may be expand such that
$$
V'(z)=-2\,\sumn_{k\geq -1}\, c_k\,\left[ z^k\, (z^2-4)^{\half}
\right]_+,
\eqno(3.5)
$$
where the $2m^{th}$ critical point is defined by
$c_k=-{(m+1)!\over 2^{m+1}\, (2m-1)!!}\,\delta_{k,2m}$. From (3.4),
the non-positive modes $\tilde J_k;\ k=0,-1,-2,\ldots$ commute
with each other and contain no differential operators except
for $\der /\der g_0$ in $\tilde J_0$, which may be replaced by
$-x$ in terms of $\beta^{-2}\,\der (\ln Z)/\der g_0=-x$.
Therefore these may be considered as a new coordinate system for
the space of the coupling constants instead of $c_k$'s or $g_k$'s.
These are related to $c_k$ such that
$$
\tilde J_{-k}~{\buildrel \rm def \over =}~
2\, (k+1)\, p_{k}=4\, c_k-c_{k-2},~~~~~~~(k\geq 0)
\eqno(3.6)
$$
where $c_{-2}=x-1$, since ${\rm Res}_{z=\infty}\,
\left[\sumn_{k\geq -1}\, c_k\, z^k (z^2-4)^{\half}\right] =-1$.
On the other hand, the positive modes $J_k,\ k=1,2,\ldots$, are
the pure differential operators w.r.t.\ $p_k$, namely
$$
\eqalign{
\tilde J_k&=\beta^{-2}\, \left[ {k\over 4(k+1)}
{\der\over\der p_k}-{\der\over\der p_{k-2}}\right]\, , ~~~~~~~
(k\geq 2)\cr
\tilde J_1&={1\over 8}\, \beta^{-2}\, {\der\over\der p_1}.\cr}
\eqno(3.7)
$$
This is not the case if we write down $J_k$'s as differential
operators w.r.t.\ $g_k$'s, {\it i.e.}
$$
\tilde J_k=\tilde J^+_k+\tilde J^-_k,~~~~~~~(k\geq 1)
\eqno(3.8)
$$
where
$$
\eqalign{
\tilde J^+_kZ&=-\int_{C_{\infty}}{dz\over 2\pi i}\,
(z^2-4)^{\half}\, z^{k-1}\,\beta^{-2}\,\langle\,
\tr ({1\over z-\phi})\,\rangle\, Z,\cr
&=-\beta^{-1}\,\langle\,\spc^{k-1}\,
(\spc^2-4)^{\half}\,\rangle\, Z,\cr}
\eqno(3.9a)
$$ and $$
\eqalign{
\tilde J^-_k&=\half\,\int_{C_{\infty}}{dz\over 2\pi i}\,
(z^2-4)^{\half}\, z^{k-1}\, V'(z),\cr
&={\rm linear~combination~of~}g_0,\ g_1,~\ldots ~.\cr}
\eqno(3.9b)
$$
Since $\tilde J^-_k$ is a $t$ independent number, we can absorb
$\tilde J^-_k$ into the integration constant of the trace
$\langle~~\rangle$ for each $\tilde J^+_k$, and obtain
$$
\tilde J_kZ=-\beta^{-1}\,\langle\,\spc^{k-1}\,
(\spc^2-4)^{\half}\rangle\, Z.
\eqno(3.10)
$$

Defining the scaling parameters $t_k$ by
$$
\eqalign{
4\, c_{-1}&=2\,\ep^{2m+1}\,q,\cr
4\, c_0-c_{-2}&=\ep^{2m}\, t_0,\cr
4\, c_k&=\ep^{2m-k}\, 2^{-k}\, (k+1)\, t_k,~~~~~~~(k\geq 1)\cr}
\eqno(3.11)
$$
the $2m^{th}$ critical point of the two-cut model is given by
$q=0$, $t_0=t$ and $t_k=-{2^{m+1}\, (m+1)!\over (2m+1)!!}
\delta_{k,2m}$.
As $\ep\rightarrow 0$, $c_{k-2};~k\geq 1$ in (3.6) and
$\der /\der p_k;~k\geq 2$ in (3.7) become negligible due to
the relative $\ep$ factor, and hence $\tilde J_k$ scales as
$$
\eqalign{
\tilde J_k&=-\kappa^4\,\ep^{2m+k}\,2^{k-1}\,
{\der\over\der t_{k-2}},~~~~~~~(k\geq 2)\cr
\tilde J_1&=\ep^{2m+1}\,\left(2\,\alpha
-\kappa^4\,{\der\over\der t_{-1}}\right) ,\cr
\tilde J_{-k}&=\ep^{2m-k}\, 2^{-k}\, (k+1)\, t_k, ~~~~~~~~~~
(k\geq 0)\cr}
\eqno(3.12)
$$
where from (3.9) one can obtain
$$
\eqalign{
2\,\alpha&=q-\kappa^2\,\beta\,\sumn_{l\geq 1}\,
c_{2l-1}\,{2^{l+1}\, (2l-1)!!\over (l+1)!},\cr
\kappa^2\,{\der\over\der t_{-1}}&=-\beta^{-1}\,
{\der\over\der g_1}.\cr}
\eqno(3.13)
$$
Using (3.12) in (3.10), one can find the commutative flows in
the NLS hierarchy such that
$$
\eqalign{
\kappa^2\,{\der F\over\der t_k}&=2^{-k-1}\,\ep^{-k-1}
\langle\spc^{k+1}\, (\spc^2-4)^{\half}\rangle ,\cr
&=2i\,\int_t^{\infty}dt\,\tr\bra tL^{k+1}\ket t,\cr
&=-2\,\int_t^{\infty}dt\, H_{k+1},\cr}
\eqno(3.14)
$$
for $k\geq -1$, where $F=\ln Z$ and the trace on the $2\times 2$
matrix is necessary in order to sum up the contributions from even
and odd polynomials.
Since $\der F/\der t_{-1}=0$ follows from $H_0=0$, the $t_{-1}$
derivative vanishes in $\tilde J_1$.
Note that $\alpha$ appears in these relations because it cannot
be absorbed into the integration constant.
$\alpha$ will be identified with the integration constant
of the third string equation, and proved to be independent of
$t_k$'s in $\S4.2$.

\endpage
\sub{3.2~~Virasoro Constraints}

Multiplying (3.1) once again with $J(z)$, one can find the discrete
Ward identity [12]
$$
:J(z)^2:\, Z={1\over 4}\,\Delta (z)\, Z \, ,
\eqno (3.15)
$$
where :\ : denotes the normal ordering, and $\Delta (z)$ is an
entire function defined by
$$
\Delta (z)=\left( V'(z)\right)^2-4\,\langle\,\left(
{V'(z) -V'(\spc )\over z-\spc}\right)\rangle .
\eqno(3.16)
$$
Hence $Z$ satisfies
$$
\int_{C_{\infty}}{dz\over 2\pi i}\, z^{n+1}\,
:J(z)^2:\, Z=0, ~~~~~~~(n\geq -1)
\eqno (3.17)
$$
for which one can find the discrete Virasoro constraints (1.10)
by calculating the contour integral. Alternatively (3.17) may
be rewritten as
$$
\eqalignno{
\int_{C_{\infty}}{dz\over 2\pi i}\,&:J(z)^2:\, Z=0,&(3.18a)\cr
\int_{C_{\infty}}{dz\over 2\pi i}\,&z\, :J(z)^2:\, Z=0,&(3.18b)\cr
\int_{C_{\infty}}{dz\over 2\pi i}\,&z^{n+1}\, (z^2-4)\,
:J(z)^2:\, Z=0,~~~~~~~(n\geq -1)&(3.18c)\cr}
$$
and calculating the contour integral we can replace (1.10) with
$$
\eqalignno{
\left(\sumn_{k\in\Z}\, 2^{-2k}\,\tilde J_{2k}\right)\,
\left(\sumn_{l\in\Z}\, 2^{-2l-1}\,\tilde J_{2l+1}\right)\, Z&=0,
&(3.19a)\cr
\left[\left(\sumn_{k\in\Z}\, 2^{-2k}\,\tilde J_{2k}\right)^2+\left(
\sumn_{l\in\Z}\, 2^{-2l-1}\,\tilde J_{2l+1}\right)^2\right]\, Z&=0,
&(3.19b)\cr
\left(\sumn_{k\in\Z}\, :\tilde J_{-k}\,\tilde
J_{n+k-2}:\right)\, Z&=0.
&(3.19c)\cr}
$$
{}From (3.3) and (2.6), we can get
$$
\eqalign{
\left(\sumn_{k\in\Z}\, 2^{-2k}\,\tilde J_{2k}\right)\, Z
&=\int_{C_{\infty}-C_0}{dz\over 2\pi i}\, z\, (z^2-4)^{-\half}\,
J^{even}(z)\, Z,\cr
&=\int_{C_{\infty}}{dz\over 2\pi i}\, (z^2-4)^{-\half}\,\left[
\half z\, V'(z)-1\right]\, Z,\cr}
\eqno(3.20a)
$$
$$
\eqalign{
\left(\sumn_{l\in\Z}\, 2^{-2l-1}\,\tilde J_{2l+1}\right)\, Z
&=\int_{C_{\infty}-C_0}{dz\over 2\pi i}\, (z^2-4)^{-\half}\,
J^{odd}(z)\, Z,\cr
&=\int_{C_{\infty}}{dz\over 2\pi i}\, (z^2-4)^{-\half}\,\half
V'(z)\, Z,\cr}
\eqno(3.20b)
$$
where the r.h.s.\ of both (3.20a) and (3.20b) vanish automatically,
since the critical potential (3.5) always satisfies (2.5).
Hence (3.19a) and (3.19b) are satisfied trivially before taking the
scaling limit.
In terms of the scaling of $\tilde J_k$ in the vicinity of
the $2m^{th}$ critical point, (3.19c) scales to the continuum
Virasoro constraints
$$
L_n\, Z=0,~~~~~~~(n\geq -1)
\eqno(3.21)
$$
where
$$
\eqalign{
L_{-1}&=\sumn_{k=1}^{\infty}\, (k+1)\, t_k\,{\der\over\der t_{k-1}}
-2\, \kappa^{-4}\,\alpha\, t,\cr
L_0&=\sumn_{k=0}^{\infty}\, (k+1)\, t_k\,{\der\over\der t_k}
-\kappa^{-4}\,\alpha^2,\cr
L_1&=\sumn_{k=0}^{\infty}\, (k+1)\, t_k\,{\der\over\der t_{k+1}}
+\alpha\,{\der\over\der t},\cr
L_n&=\sumn_{k=0}^{\infty}\, (k+1)\, t_k\,{\der\over\der t_{k+n}}
-{\kappa^4\over 4}\,\sumn_{k=1}^{n-1}\,{\der^2\over\der t_{k-1}\,
\der t_{n-k-1}}+\alpha\,{\der\over\der t_{n-1}}.\cr}
\eqno(3.22)
$$
These constraints have been found already in refs.\ [17,18] by
integrating the string equations
$$
\eqalign{
\sumn_{k\geq 0}\, (k+1)\, t_k\, F_k&=0,\cr
\sumn_{k\geq 0}\, (k+1)\, t_k\, G_k&=0.\cr}
\eqno(3.23)
$$
Finally, defining $\av {\sigma_k}=\int^tdt\, H_{k+1}$, we arrive at
$$
\eqalign{
\sumn_{k=1}^{\infty}\, (k+1)\, t_k\,\av {\sigma_{k-1}}
-\kappa^{-2}\,\alpha\, t&=0,\cr
\sumn_{k=0}^{\infty}\, (k+1)\, t_k\,\av {\sigma_k}
-\kappa^{-2}\,{\alpha^2\over 2}&=0,\cr
\sumn_{k=0}^{\infty}\, (k+1)\, t_k\,\av {\sigma_{k+1}}
+\alpha\,\av {\sigma_0}&=0,\cr
\sumn_{k=0}^{\infty}\, (k+1)\, t_k\,\av {\sigma_{k+n}}
+\alpha\,\av {\sigma_{n-1}}&\cr
-{\kappa^2\over 2}\,\sumn_{k=1}^{n-1}\,
(\av {\sigma_{k-1}\,\sigma_{n-k-1}}&
+\av {\sigma_{k-1}}\,\av {\sigma_{n-k-1}})=0.\cr}
\eqno(3.24)
$$

\beginsection 4. NLS Hierarchy

\pub{4.1~~Flows and String Equations}

Gel'fand and Dikii [20] have obtained the Lax pair for the
general matrix differential operator $L$ in terms of the diagonal
kernel $R(z)=\bra t(L-z\,E)^{-1}\ket t$. For our first order
$2\times 2$ matrix operator $L$, their theorem 7 can be read as
$$
\sumn_{k\geq 0}\, P_k\,z^{-k-1} ~{\buildrel \rm def \over =}~
R(z)\,\sigma_2\, (L-z\, E)^{-1}=(L-z\, E)^{-1}\,\sigma_2\, R(z),
\eqno(4.1)
$$
where the last equality follows from their theorem 6.
First few of $P_k$ are
$$
\eqalign{
P_0&=-{i\over 2}\,\sigma_2,~~~~~~~~~~P_1=-\half\,D,\cr
P_2&={i\over 2}\,\sigma_2\, [\, D^2
-\half\,\{ D,f\sigma_3+g\sigma_1\}
-\half\, (f\sigma_3+g\sigma_1)^2\, ],\cr
P_3&=\half\, [\, D^3
-{3\over 4}\,\{ D^2,f\sigma_3+g\sigma_1\}
-{3\over 4}\,\{ D,(f\sigma_3+g\sigma_1)^2\}\, ],\cr}
\eqno(4.2)
$$
and the $k^{th}$ equation in the NLS hierarchy may be defined as
$$
{\der L\over\der t_k}=2[L,P_{k+1}]
=2\, G_{k+1}\,\sigma_3-2\, F_{k+1}\,\sigma_1.
\eqno(4.3)
$$
Here the $k^{th}$ NLS flows
$$
\eqalign{
{\der f\over\der t_k}&=2\, F_{k+1},\cr
{\der g\over\der t_k}&=2\, G_{k+1},\cr}
\eqno(4.4)
$$
are all commutative for $k\geq -1$.

The original definition in ref.\ [20] has no flows for $k=-1$
due to the extra $(k+1)$ factor appearing in the r.h.s.\ of (4.4),
which we have been normalized to the definition of $t_k$.
Both are consistent with the previous definition based on the
mapping from the Toda flows. In fact, by taking the $t$
derivative twice on (3.14), one can find that
$$
{\der (f^2+g^2)\over\der t_k}=2\, H'_{k+1}=
4\, (f\, F_{k+1}+g\, G_{k+1}),
\eqno(4.5)
$$
for $k\geq -1$, which agrees with (4.4), while the $(-1)^{th}$
flow vanishes on $f^2+g^2$ anyway. Actually these two definitions
are equivalent due to the factorization of the $t_{-1}$ dependence.
This can be done as follows. First introduce the complex notation
$\psi =f+i\, g$ and $U_k=F_k+i\, G_k~~(k\geq -1)$; thus the
$t_{-1}$ dependence of $\psi$ can be factorized as $\psi =
e^{-2it_{-1}}\,\psi_0$ by solving the $(-1)^{th}$ equation
$$
\eqalign{
{\der f\over\der t_{-1}}&=2\, g,\cr
{\der g\over\der t_{-1}}&=-2\, f.\cr}
\eqno(4.6)
$$
Rewriting (2.19) with the complex notation\footnote{$^{\star}$}
{\small The first equation in the NLS hierarchy is
$i\,\psi_{t_1}=-{\psi}''/2+|\psi |^2\,\psi$, the nonlinear
Schr\"odinger equation.}
$$
\eqalign{
U_{k+1}&=-i\,\psi\, H_k+{i\over 2}\, U'_k,\cr
H'_k&=\psi^{*}\, U_k+\psi\, U_k^{*},\cr}
\eqno(4.7)
$$
and using the induction w.r.t.\ $k$ starting with $U_{-1}=-i\,\psi$,
one can find that all $U_k$ have the same $t_{-1}$ dependence as
$\psi$, namely $U_k=e^{-2it_{-1}}\, (U_k)_0$, and $t_{-1}$
independent $\psi_0$ and $(U_k)_0$ again satisfy the same
recursion relations as (4.7), where $H_k$'s are $t_{-1}$
independent by virtue of (4.5).
Therefore we can conclude that the $t_{-1}$ dependence is
redundant, and $\psi$ has a global phase ambiguity.
We will leave the $(-1)^{th}$ flow non-vanishing for convenience.

It is noteworthy that the string equations (3.23)
can be rewritten as
$$
[M,L]=E,
\eqno(4.8)
$$
by defining $M=\sumn_{k\geq 0}\, (k+1)\, t_k\, P_k$.
This is the scaling from (1.8) instead of (2.1).

\sub{4.2~~Invariance of the String Equations}

We will prove the invariance of the string equations under the
NLS flow. Defining
$$
K_{-1}=\sumn_{k\geq 0}\, (k+1)\, t_k\,{\der\over\der t_{k-1}},
\eqno(4.9)
$$
the string equations (3.23) may be expressed as
$$
K_{-1}\, f=K_{-1}\, g=0,
\eqno(4.10)
$$
and integrating $K_{-1}\, (f^2+g^2)=0$ once w.r.t.\ $t$, one can
obtain the third string equation
$$
\sumn_{k\geq 0}\, (k+1)\, t_k\, H_k=c,
\eqno(4.11)
$$
where the integration constant $c$ may not vanish, and by taking
the $t$ derivative once on $L_{-1}\, Z=0$, one can identify it
as $c=\kappa^{-2}\,\alpha$.

The string equations (4.10) and (4.11) are invariant
under the NLS flows if
$$
\eqalignno{
(l+1)\, F_l+K_{-1}\, F_{l+1}&=0,&(4.12a)\cr
(l+1)\, G_l+K_{-1}\, G_{l+1}&=0,&(4.12b)\cr
(l+1)\, H_l+K_{-1}\, H_{l+1}&=0,&(4.12c)\cr}
$$
hold for $l\geq -1$. Here we have used the $t_k$ independence of
$\alpha$ to get (4.12c) from (4.12a) and (4.12b).
$\der\alpha /\der t_{-1}=\der\alpha /\der t_0=0$ follows
directly from the string equations, and $\der\alpha /\der
t_k=0~~(k:{\rm even})$ follows from the definition of $\alpha$,
while $\der\alpha /\der t_k=0~~(k:{\rm odd})$ is a result of the
fine-tuning of the odd couplings to make $\alpha$ finite. Now we can
prove (4.12) by the induction w.r.t.~$l$ starting from $l=-1$, which
is nothing other than the string equations (4.10). Suppose that
(4.12) holds for $l\leq k-1$, then the recursion relations yield
$$
\eqalign{
K_{-1}\, F_{k+1}&=g\, K_{-1}\, H_k-\half\, (K^{-1}\, G_k)'+
\half\, {\der\over\der t_{-1}}\, G_k,\cr
&=-(k+1)\, F_k.\cr}
\eqno(4.13)
$$
(4.12b) may be proved similarly.

In the presence of (4.11), we can derive alternative string
equations by taking the $t$ derivative once on (4.10); namely
$$
K_0\, f=-f,~~~~~~~~~~K_0\, g=-g,
\eqno(4.14)
$$
where
$$
K_0=\sumn_{k\geq 0}\, (k+1)\, t_k\,{\der\over\der t_k}~-~
c\, {\der\over\der t_{-1}}.
\eqno(4.15)
$$
Taking the $t$ derivative $n$ times on (4.14), one can find that
$$
\eqalign{
K_0\, f^{(n)}&=-(n+1)\, f^{(n)},\cr
K_0\, g^{(n)}&=-(n+1)\, g^{(n)}.\cr}
\eqno(4.16)
$$
It is convenient to define ``dimensions" by $d(f)=d(g)=-1$ and
$d(t_k)=k+1$.
Since $K_0$ is a first order differential operator, {\it i.e.}
$K_0\,(AB)=(K_0\, A)\, B+A\, (K_0\, B)$, is therefore an operator
counting the dimensions of the differential polynomial $A[f,g]$
if $f$ and $g$ are the solution of (4.14); namely
$$
K_0\, A[f,g]=d(A)\, A[f,g],
\eqno(4.17)
$$
where $A[f,g]$ may have explicit $t_k$ dependences except
for $t_{-1}$.
It is then easy to prove the invariance of (4.14)
under the NLS flows directly.

By restricting the potential to be an even function of $\lam$
($b\neq 0$), we can find a constraint $g=0$, and consequently the
NLS hierarchy is reduced to the mKdV hierarchy with $t_{2k-1}=0$
for $k\geq 0$, where the odd flows are all frozen because they
vary the constraint $g=0$.
Only the second string equation $K_{-1}\, g=0$ remains nontrivial
in this reduction, and one can obtain the mKdV flows and the string
equation
$$
\eqalignno{
{\der f\over\der t_{2k}}=&S'_k[f],~~~~~~~(k\geq 0)&(4.18)\cr
\sumn_{k\geq 0}\, (2k+1)\,&t_{2k}\, S_k[f]=0,&(4.19)\cr}
$$
where the string susceptibility is given by $\der_t^2\,
F=\kappa^{-2}\, f^2$. It is more appropriate to describe (4.19)
as the once integrated form of $K_0\, f=-f$, which is manifestly
invariant under the mKdV flows.

\beginsection 5. ZS Hierarchy

\pub{5.1~~Rotation from NLS to ZS}

One can deal with the anti-hermitian matrix model in the same
manner as we did with the hermitian matrix model in $\S2$.
Hollowood et al.~[18] have already studied the anti-hermitian
matrix model with real couplings, and found what they call the
Zakharov-Shabat (ZS) hierarchy. Here we will re-investigate this
model in terms of the mapping we have found.
By rotating the anti-hermitian matrix $\tilde\phi$ to the hermitian
matrix $\phi =i\,\tilde\phi$, one can find that the hermitian matrix
model has a complex potential, namely $g_k=(-i)^k\,\tilde g_k$,
where $\tilde g_k$'s are the couplings of the anti-hermitian model,
and all real valued.
Since the eigenvalues of $\tilde\phi$ distributes along the
imaginary axis, the limit values of $R_n$ and $S_n$ also rotate
as $\tilde a=-i\, a$ and $\tilde b=-i\, b$, where $a$ and
$b$ are the limits defined previously in the hermitian model,
and consequently the scaling functions are given by $\tilde f=-f$
and $\tilde g=-i\, g$ with $f$ and $g$ being real valued.
Defining the coupling constant of the hermitian matrix model by
$c_k=(-i)^{2+k}\,\tilde c_k~~(\tilde c_k\in\R )$, and the spectral
parameter by $z=i\,\tilde z~~(z\in\R )$, one can find that
$$
1-\half\,\tilde z\, V'(\tilde z)=\sumn_{k\geq -1}\,
c_k\,\left[ (z-b)^{k+1}\, (z-\lam_+)^{\half}\,
(z-\lam_-)^{\half}\right]_+,
\eqno(5.1)
$$
in which the $k^{th}$ critical potential is either real or pure
imaginary depending on $k$ even or odd as we expect.
$c_{2k}$ must satisfy the constraint $\sumn_{k\geq 1}\,
(2\, a^2)^{k+1}\,{(2k-1)!!\over (k+1)!}\, c_{2k}$ $=-1$,
but unlike the previous model, $c_{2k-1}$ does not have to be
zero, because the pure imaginary potential does not contribute
to the eigenvalue density. Now the Lax operator
$$
\tilde L=-i\, L
=-i\, \sigma_2\, D+\sigma_1\, f-i\,\sigma_3\, g,
\eqno(5.2)
$$
turns out to be complex, thus we need an $SU(2)$ rotation in order
to make it pure imaginary, {\it i.e.}
$$
L=\sigma_3\, D+i\,\sigma_2\, f+\sigma_1\, g.
\eqno(5.3)
$$

Under the rotation from $c_k$ to $(-i)^{k+2}\, \tilde c_k$,
the NLS hierarchy is rotated to the ZS hierarchy by
$t_k\rightarrow i^k\,\tilde t_k$, where we have changed
$\tilde t_k$ to $(-1)^{k+1}\,\tilde t_k$ for convenience.
Note that $\tilde t_0=t$, and the $2m^{th}$ critical point is
given by
$\tilde t_k={(-2)^{m+1}\, (m+1)!\over (2m+1)!!}\,\delta_{k,2m}$.
Defining new differential polynomials in such a way that the flow
equations appear to be the same, namely
$$
F_k\rightarrow (-i)^{k+1}\, \tilde F_k, ~~~~~~~
G_k\rightarrow (-i)^k\, \tilde G_k, ~~~~~~~
H_k\rightarrow (-i)^{k+1}\, \tilde H_k,
\eqno(5.4)
$$
with $\tilde F_{-1}=\tilde G_{-1}=0$, $\tilde H_{-1}=1$,
$\tilde F_0=g$, $\tilde G_0=f$, $\tilde H_0=0$;
we can find the residue
$$
\bra tL^k\ket t=\half\,\left( \tilde F_k\,\sigma_1+
i\,\tilde G_k\,\sigma_2+\tilde H_k\,E\,\right) ,
\eqno(5.5)
$$
with the recursion relations
$$
\eqalign{
\tilde F_{k+1}&=g\,\tilde H_k+\half\,\tilde G'_k,\cr
\tilde G_{k+1}&=f\,\tilde H_k+\half\, \tilde F'_k,\cr
\half\,\tilde H'_k&=g\,\tilde G_k-f\,\tilde F_k.\cr}
\eqno(5.6)
$$
Now all negative signs in the Virasoro constraint (3.22) change
to the positive signs because of the $i^k$ factor, but (3.24)
does not change because
$$
\kappa^2\, {\der F\over\der \tilde t_k}=
2\,\int_t^{\infty}dt\,\tilde H_{k+1}.
\eqno(5.7)
$$
In terms of $\psi =f-g$, $\bar\psi =f+g$, $U_k=\tilde F_k-
\tilde G_k$, and $\bar U_k=\tilde F_k+\tilde G_k$, the ZS
flows and the string equations may be rewritten as
$$
{\der\psi\over\der \tilde t_k}=2\, U_{k+1},~~~~~~~
{\der\bar\psi\over\der \tilde t_k}=2\,\bar U_{k+1},
\eqno(5.8)
$$
$$
\eqalign{
\sumn_{k\geq 0}\, (k+1)\, \tilde t_k\, U_k=0,~~~~&~~~
\sumn_{k\geq 0}\, (k+1)\, \tilde t_k\,\bar U_k=0,\cr
\sumn_{k\geq 0}\, (k+&1)\,\tilde t_k\,\tilde H_k=\tilde c.\cr}
\eqno(5.9)
$$

For the even potential, again $g=0$ reduces the ZS hierarchy
to the mKdV hierarchy, for which the differential polynomials are
$F_{2k+1}=(-1)^k\,\half\, S'_k[f]$, $G_{2k}=(-1)^k\, S_k[f]$,
$H_{2k+1}=(-1)^k\, (\half\, S'_k[f]-R_{k+1}[f^2+f'])$, and zero
otherwise, hence we can find the same mKdV flows as (4.18) and
the same string equation as (4.19) by assigning $t_{2k}=(-1)^k\,
\tilde t_{2k}$, and also the string susceptibility is again
$\der_t^2\, F=\kappa^{-2}\, f^2$. Note that the $2m^{th}$ critical
point, $t_k=-{2^{m+1}\, (m+1)!\over (2m+1)!!}\,\delta_{k,2m}$,
is precisely same as before.

Hollowood et al.~[18] have found an alternative way to freeze
odd flows, namely $\bar\psi =e^{2\tilde t_{-1}}$ and
$u=\psi\bar\psi$, for which one can find that
$$
\eqalign{
U_{2k}&=(-1)^k\, (\half\, D^2-u)\, R_k[u],\cr
U_{2k+1}&=2\, H_{2k+2}=(-1)^k\, R'_{k+1}[u],\cr}~~~~~~~
\eqalign{
\bar U_{2k}&=H_{2k-1}=(-1)^k\, R_k[u],\cr
\bar U_{2k+1}&=0,\cr}
\eqno(5.10)
$$
with $R_k[u]$ being the $k^{th}$ Gel'fand-Dikii differential
polynomials of the KdV hierarchy, therefore this is the reduction
into the KdV hierarchy
$$
\eqalignno{
{\der u\over\der t_{2k}}=&2\, R'_{k+1}[u],&(5.11)\cr
\sumn_{k\geq 0}\, (2k+1)\,&t_{2k}\, R_k[u]=0,&(5.12)\cr}
$$
where the string susceptibility is given by $\der_t^2\, F=
\kappa^{-2}\, u$. This result agrees with the one-cut family of
the hermitian models (renormalization of $t_{2k}$ is necessary
to adjust the critical points). The KdV string equation is the
once integrated form of the third string equation with $\tilde c
=\half$, and is therefore manifestly invariant under the KdV flows.

\sub{5.2~~Quantization of $\tilde c$}

The NLS hierarchy may be obtained from the two component KP
hierarchy by the so-called reduction procedure, or alternatively,
{}from the hierarchy of soliton equations associated with the
$A_1^{(1)}$ Kac-Moody algebra, for which two different vertex
operator realizations (the principal and the homogeneous pictures)
respectively lead to the KdV and the NLS hierarchies [25].
In the homogeneous picture, the vertex operator has one free
parameter $\gamma$ due to the zero mode of the scalar field,
where $\gamma$ could be any element of the finite root lattice,
and consequently the tau-function depends also on $\gamma$.
For $A_1^{(1)}$, one finds $\gamma =m\,\alpha_1~~(m\in\Z )$, and
the scaling functions $\psi$ and $\bar\psi$ may be assigned to be
$$
\psi =\kappa^{2+4\tilde c}\,{\tau_{m+1}\over\tau_m},~~~~~
\bar\psi =\kappa^{2-4\tilde c}\,{\tau_{m-1}\over\tau_m},~~~~~
\psi\,\bar\psi =-\kappa^4\,\der_t^2\,\ln\tau_m,
\eqno(5.13)
$$
for which one can get the NLS and the ZS hierarchies by taking
particular real slices, {\it i.e.} $\bar\psi =\psi^{*}$ (resp.\
$\psi ,~\bar\psi\in\R$).

By following this tau-function formalism, Hollowood et al.\ [18]
have found that the Virasoro constraints
$L_n(\tilde c)\,\tau_m=0~~(n\geq -1)$ produce
$$
L_n(\tilde c\,\mp 1)\,\tau_{m\pm 1}=0,~~~~~~~~(n\geq -1)
\eqno(5.14)
$$
which strongly suggests the quantization of $\tilde c$.

Actually $\tilde c$ must be quantized as $\tilde c\in\Z /2$ due to
the following mechanism. If we rescale $\ep$ to $a\,\ep$, while
keeping the model and $\kappa$ fixed, $\tilde t_k$ changes to
$a^{k+1}\,\tilde t_k$. This can be seen most easily in the first
line of (3.14), where $F$ and $\av{~~}$ do not change by assumption.
Since (4.14) holds as well in the ZS hierarchy, and the
$\tilde t_{-1}$ dependence may be factorized as $\psi
=e^{-2\tilde t_{-1}}\,\psi_0$ and $\bar\psi =e^{2\tilde t_{-1}}\,
\bar\psi_0$, $\psi$ (and $\bar\psi$) has the anomalous dimension
$-2\tilde c$ (resp.\ $2\tilde c$), {\it i.e.}
$$
\eqalign{
\sumn_{k\geq 0}\, (k+1)\,\tilde t_k\,{\der\psi\over\der
\tilde t_k}&=-(1+2\tilde c)\,\psi ,\cr
\sumn_{k\geq 0}\, (k+1)\,\tilde t_k\,{\der\bar\psi\over\der\tilde
t_k}&=-(1-2\tilde c)\,\bar\psi .\cr}
\eqno(5.15)
$$
Therefore if $2\tilde c$ is not integral, the continuum limit
$\ep\rightarrow 0$ becomes non-analytic with respect to $\ep$,
and hence $\tilde c$ must be half-integral.

(5.15) also explains the $\kappa$ factors appearing in (5.13),
because the rescaling of $\tilde t_0$ may be realized by the
rescaling of $\kappa^2$. If the way $R_n$ and $S_n$ approach
their limit values is scale invariant, $\tilde c$ must be zero,
which is the case for the reduction to the mKdV hierarchy.
On the other hand, the reduction to the KdV hierarchy has been
achieved by the constraint $\bar\psi =e^{2\tilde t_{-1}}$, thus
the scaling dimension of $\bar\psi$ must be zero, and hence
(5.15) gives $\tilde c=1/2$ which agrees with the previous
calculation.

We have derived (4.14) from the string equations (4.10) and (4.11),
therefore any solution of the string equations (5.9) satisfies
(5.15) for arbitrary value of $\tilde c$. However, there is no
guarantee that (5.15) with arbitrary $\tilde c$ is compatible with
the constraint we require such as $\tilde t_k=0~~(k:{\rm odd})$.
Above examples show that the compatibility actually breaks down.
But of course, not all constraints quantize $\tilde c$, for
instance, in the topological phase emerging at $\tilde t_1\neq 0$
and $\tilde t_k=0$ for $k\geq 2$, the string equations have exact
solution such as
$$
\psi =-\tilde c^{\; r}\,\tilde t_1^{~-\half -\tilde c}\,
\exp (-2\tilde t_{-1}+{t^2\over 2\tilde t_1}),~~~~~
\bar\psi =\tilde c^{\; s}\,\tilde t_1^{~-\half +\tilde c}\,
\exp (2\tilde t_{-1}-{t^2\over 2\tilde t_1}),
\eqno(5.16)
$$
with $r+s=1$, which satisfy (5.15) and the third string equation
for arbitrary $\tilde c$.

\beginsection 6. Discussions and Conclusion

\pub{6.1~~Hermitian Models and the NLS Hierarchy}

The $2m^{th}$ mapping we have obtained in the hermitian model is
$$
\eqalign{
2\, c_{-1}-\sumn_{l\geq 1}\,{2^{l+1}\, (2l-1)!!\over (l+1)!}
\, c_{2l-1}&=\ep^{2m+1}\, 2\,\alpha ,\cr
4\, c_0-c_{-2}&=\ep^{2m}\, t_0,~~~~~~~~~~~~~~~~~~~~~~~~\; (t_0=t)\cr
4\, c_k&=\ep^{2m-k}\, 2^{-k}\, (k+1)\, t_k,~~~~~~~(k\geq 1)\cr}
\eqno(6.1)
$$
which maps the $2m^{th}$ critical potentials, {\it i.e.} $c_{2m}<0$
and $c_k=0~~(k<2m)$, to the subspace $t_{2m}<0$ and $t_k=0~~(k>2m)$.
At $\ep =0$, only the $2m^{th}$ critical point is admissible,
because $c_k=0~~(k<2m)$ and $t_k=0~~(k>2m)$ respectively imply
$t_k=0~~(k<2m)$ and $c_k=0~~(k>2m)$ at $\ep =0$, thus $t_{2m}=
-{2^{m+1}\, (m+1)!\over (2m+1)!!}$ from the constraint. Owing to
the positivity of the density of the eigenvalue distribution,
$c_k$ must vanish for $k$ odd, so that no odd critical points are
allowed within the hermitian model; nevertheless we may turn on
$\alpha,t_1,\ldots ,t_{2m-1}$, since $c_{-1}=\ldots =c_{2m-1}=0$
hold as $\ep\rightarrow 0$.
To this extent, we can perturb the system away from the $2m^{th}$
critical point and generalize the system to allow $c_{2m}>0$ as well
as the odd critical points. (6.1) should be understood in this sense,
and also we extend it to the $(2m-1)^{th}$ mapping, where any point
of the subspace $t_{2m-1}\neq 0$ and $t_k=0$ $(k\neq 0,2m-1)$ may be
chosen as the $(2m-1)^{th}$ critical point.

Among these ``time" parameters, the $(-1)^{th}$ parameters $\alpha$
and $t_{-1}$ are very special. $\alpha$ may be identified with the
integration constant of the third string equation, and in order to
get finite $\alpha$ we must fine-tune the odd couplings such that
$\sumn_{k\geq 1}\,{2^l\, (2k-1)!!\over (l+1)!}\, c_{2k-1}=c_{-1}$,
or equivalently, $2\,\alpha+\sumn_{k\geq 1}\,\ep^{-2k}\,{k\,
(2k-1)!!\over 2^{k-1}\, (k+1)!}\, t_{2k-1}=q$; however we do not
have any flows associated with $\alpha$, instead the $(-1)^{th}$
flow changes the parameter $t_{-1}$, and its numerical value is
independent of any of the coupling constants.
The reason of the quantization of $c=\kappa^{-2}\,\alpha$
is therefore not because of the momentum zero-mode quantization such
as $2\, c=-\kappa^2\,\der /\der t_{-1}$, but because of the
analyticity of the continuum limit discussed in $\S5.2$.
This is somehow analogous to the quantization of the central
extension $c$ in the Kac-Moody algebra, where $c$ commutes with
its dual $d$ (the zero mode of the Virasoro algebra).

The image of the generalized $m^{th}$ critical potentials consists
of two disconnected components depending on $t_m$ negative or
positive, which we call the negative (positive) side of the $m^{th}$
leaf. The NLS flows from the first to the $m^{th}$ order define
smooth coordinates on each side of the $m^{th}$ leaf, where the
even critical point is located on the negative side of the even
leaf, so that the odd leaves and the positive side of even leaves
are unreachable by the matrix model. The $2m^{th}$ flow cannot
pass through $t_{2m}=0$ due to the constraint we have, and because
of that, if we impose the ``physical" conditions to the solution of
the string equations, {\it i.e.} real and pole-free along the
positive real axis and proper asymptotic behavior for
$t\rightarrow\infty$, that solution has an instability at $t_{2m}=0$.

\sub{6.2~~Anti-Hermitian Models and the ZS Hierarchy}

For the anti-hermitian model, (6.1) may be modified as
$$
\eqalign{
2\, \tilde c_{-1}-\sumn_{l\geq 1}\,{(-2)^{l+1}\,
(2l-1)!!\over (l+1)!}\,\tilde c_{2l-1}&=
\ep^{m+1}\, 2\,\tilde\alpha ,\cr
-4\, \tilde c_0-\tilde c_{-2}&=\ep^m\,
\tilde t_0,~~~~~~~~~~~~~~~~~~~~~~~~\; (\tilde t_0=t)\cr
(-1)^{k+1}\, 4\,\tilde c_k&=\ep^{m-k}\, 2^{-k}\, (k+1)\,
\tilde t_k.~~~~~~~(k\geq 1)\cr}
\eqno(6.2)
$$
Rotating NLS to ZS, we can find that the above arguments hold
perfectly good, except for the fact that the anti-hermitian model
can realize the odd critical points as well, so the odd leaves
are reachable by the anti-hermitian matrix models.
Consequently, at the $(2m-1)^{th}$ critical point, $\tilde c_k<0$
for the first non-vanishing even coupling constant and $k$ must
be equal to or higher than $2m$. Furthermore it is not possible to
have only one odd coupling constant $\tilde c_{2m-1}$ non-vanishing
in order to get finite $\tilde\alpha$.

\sub{6.3~~Conclusion}

We will conclude with a few comments. So far we have studied the
mappings individually, but since these leaves cover the entire
t-space, and also the structure of the NLS (ZS) flows is universal,
we may sew them and get a universal mapping.
This universal system is however highly mathematical unless one
imposes the physical conditions, and among the points of the
t-space, only the $2m^{th}$ critical point has a definite scaling
dimension $\gamma_{\rm str}=-1/m$. In fact, by solving the
$2m^{th}$ string equation asymptotically, one can find that
$$
F(t)=\sumn_{h\geq 0}\, (\kappa^2\,
t^{\gamma_{\rm str}-2})^{h-1}\, F_h,
\eqno(6.3)
$$
which agrees with the scaling behavior of the continuum gravity
coupled with conformal matter.

The quantization of $\tilde c$ is necessary to make the continuum
limit analytic, and it can be realized by fixing the odd flows
by $g=0$, or $\bar\psi =e^{2\tilde t_{-1}}$ with $\tilde c=0$
(resp.\ $\tilde c=\half$).
Combining these results with (5.14), it seems that there are no
other inequivalent fixings to realize the half-integer $\tilde c$,
but this is an open problem, as is the interpretation of the
quantization of $\tilde c$ topologically or field theoretically
in the continuum quantum gravity.

The first critical point in the ZS hierarchy has been studied in
ref.\ [17] as the topological phase, but so far no topological
models have been found to be equivalent to this phase.
Nonetheless this does not discourage us to search for new
topological gravity, which may usher in a new geometry such
as the topological phase of the one-cut hermitian model did in
the intersection theory on the moduli space of Riemann surfaces
[14,15].

\vskip 32pt
{\parindent=0pt
{\bf Acknowledgements}

I would like to thank Prof.\ L.~N.~Chang for careful reading
of this manuscript and stimulating discussions.}

\vskip 60pt

\beginsection References

\item{1.}{D.J. Gross and A.A. Migdal, Phys.\ Rev.\ Lett.\
{\bf 64} (1990) 127; Nucl.\ Phys.\ {\bf B340} (1990) 333;
E. Br\'ezin and V.A. Kazakov, Phys.\ Lett.\ {\bf B236}
(1990) 144; M.R. Douglas and S.H. Shenker, Nucl.\ Phys.\
{\bf B335} (1990) 635.}
\item{2.}{M. Bershadsky and I.R. Klebanov, Phys.\ Rev.\ Lett.\
{\bf 65} (1990) 3088.}
\item{3.}{M. Adler, Inv.\ Math.\ {\bf 50} (1979) 219;
E. Martinec, Commun.\ Math.\ Phys.\ {\bf 138} (1991) 437.}
\item{4.}{H. Neuberger, ``Matrix Models and KdV," preprint
RU-90-58 (1990).}
\item{5.}{A. Mironov and A. Morozov, Phys.\ Lett.\ {\bf B252}
(1990) 47; H. Itoyama and Y. Matsuo, Phys.\ Lett.\ {\bf B255}
(1991) 202.}
\item{6.}{M. Douglas, Phys.\ Lett.\ {\bf B238} (1990) 176.}
\item{7.}{V.G. Dringel'd and V.V. Sokolov, J.\ Sov.\ Math.\
{\bf 30} (1985) 1975.}
\item{8.}{E. Witten, Nucl.\ Phys.\ {\bf B340} (1990) 281;
R. Dijkgraaf and E. Witten, Nucl.\ Phys.\ {\bf B342} (1990) 486;
R. Dijkgraaf, H. Verlinde and E. Verlinde, Nucl.\ Phys.\
{\bf B348} (1991) 457;
``Notes on Topological String Theory and 2D Quantum Gravity,"
preprint PUPT-1217 (1990), to appear in Proc.\ of the Cargese
Workshop.}
\item{9.}{R. Dijkgraaf, H. Verlinde and E. Verlinde,
Nucl\ .Phys.\ {\bf B348} (1991) 435.}
\item{10.}{J. Goeree, Nucl.\ Phys.\ {\bf B358} (1991) 737.}
\item{11.}{M. Fukuma, H. Kawai and R. Nakayama,
Int.\ J.\ Mod.\ Phys.\ {\bf A6} (1991) 1385;
``Infinite Dimensional Grassmanian Structure of Two-Dimensional
Quantuam Gravity," preprint UT-572-TOKYO (1990).}
\item{12.}{W. Ogura, Mod.\ Phys.\ Lett. {\bf A6} (1991) 811-818.}
\item{13.}{K. Li, Nucl.\ Phys.\ {\bf B354} (1991) 711, ibid.\ 725.}
\item{14.}{M.L. Kontsevich, Funk.\ Anal.\ Appl.\ {\bf 25} (1991)
123.}
\item{15.}{E. Witten, ``On the Kontsevich Model and Other Models
of Two Dimensional Gravity," preprint IASSNS-HEP-91/24 (1991);
S. Kharchev, A. Marshakov, A. Mirnov, A. Morozov and A. Zabrodin,
``Unification of All String Models with $c<1$," preprint ITEP-M-8/91
(1991); D. Gross and M.J. Newman, ``Unitary and Hermitian Matrices
in an External Field II: The Kontsevich Model and Continuum Virasoro
Constraints," preprint PUPT-1282 (1991).}
\item{16.}{P. Petropoulos, Phys.\ Lett.\ {\bf B261} (1991) 402;
S. Dalley, ``Critical Conditions for Matrix Models of String
Theory," preprint SHEP-90/91-6 (1990).}
\item{17.}{\v C.~Crnkovi\'c and G.~Moore, Phys.~Lett.~{\bf B257}
(1991) 322; \v C.~Crnkovi\'c, M. Douglas and G.~Moore, ``Loop
equations and the Topological Phase of Multi-cut Matrix
Models," preprint YCTP-P25-91 (1991).}
\item{18.}{C. Nappi, Mod.\ Phys.\ Lett.\ {\bf A5} (1990) 2773;
T. Hollowood, L. Miramontes, A. Pasquinucci and C. Nappi,
``Hermitian vs.\ Anti-Hermitian 1-Matrix Models and Their
Hierarchies," preprint IASSNS-HEP-91/59.}
\item{19.}{V.E. Zakharov and A.B. Shabat, Func.\ Anal.\ Appl.\
{\bf 8} (1974) 226;\br I.M. Krichever, Func.\ Anal.\ Appl.\
{\bf 11} (1977) 12.}
\item{20.}{I.M.\ Gel'fand and L.A.\ Dikii, Func.\ Anal.\ Appl.\
{\bf 11} (1977) 93.}
\item{21.}{For instance, it has been shown that the physical
solution is unstable under the flow from the second critical
point to the pure gravity (the first critical point) in the
one-cut family of the hermitian models:
M.R. Douglas, N. Seiberg and S.H. Shenker, Phys.\ Lett.\
{\bf B244} (1990) 381.}
\item{22.}{H. Neuberger, Nucl.\ Phys.\
{\bf B352} (1991) 689.}
\item{23.}{V. Periwal and D. Shevitz, Phys.\ Rev.\ Lett.\
{\bf 64} (1990) 1326; Nucl.\ Phys.\ {\bf B344} (1990) 731.}
\item{24.}{\v C.~Crnkovi\'c, M. Douglas and G.~Moore, Nucl.\
Phys.\ {\bf B360} (1990) 507.}
\item{25.}{V.G. Kac and M. Wakimoto, Proc.\ of Symposia in Pure
Math., {\bf 49} (1989) 191.}
\bye
\end